\documentclass[fleqn,usenatbib]{mnras}
\usepackage{amsmath}
\usepackage{graphicx}
\usepackage{amssymb,txfonts}
\usepackage{xspace}

\newbox\grsign \setbox\grsign=\hbox{$>$} \newdimen\grdimen \grdimen=\ht\grsign
\newbox\simpropbox
\setbox\simpropbox=\hbox{\raise.5ex\hbox{$\propto$}\llap
 {\lower.5ex\hbox{$\sim$}}}\ht2=\grdimen\dp2=0pt

\title[The lamppost model]{The lamppost model: effects of photon trapping, the bottom lamp and disc truncation}
\author[A. Nied{\'z}wiecki and A. A. Zdziarski]
{Andrzej Nied{\'z}wiecki$^1$\thanks{E-mail: niedzwiecki@uni.lodz.pl (AN), aaz@camk.edu.pl (AAZ).} and Andrzej A. Zdziarski$^2$\footnotemark[1]\\ 
$^1$Department of Physics, {\L}{\'o}d{\'z} University, Pomorska 149/153, 90-236 {\L}{\'o}d{\'z}, Poland\\
$^2$Centrum Astronomiczne im.\ M. Kopernika, Bartycka 18, PL-00-716 Warszawa, Poland\\
}

\pagerange{\pageref{firstpage}--\pageref{lastpage}}
\pubyear{2018}

\begin{document}
\maketitle

\label{firstpage}

\begin{abstract}
We study the lamppost model, in which the primary X-ray sources in accreting black-hole systems are located symmetrically on the rotation axis on both sides of the black hole surrounded by an accretion disc. We show the importance of the emission of the source on the opposite side to the observer. Due to gravitational light bending, its emission can increase the direct (i.e., not re-emitted by the disc) flux by as much as an order of magnitude. This happens for near to face-on observers when the disc is even moderately truncated. For truncated discs, we also consider effects of emission of the top source gravitationally bent around the black hole. We also present results for the attenuation of the observed radiation with respect to that emitted by the lamppost as functions of the lamppost height, black-hole spin and the degree of disc truncation. This attenuation, which is due to the time dilation, gravitational redshift and the loss of photons crossing the black-hole horizon, can be as severe as by several orders of magnitude for low lamppost heights. We also consider the contribution to the observed flux due to re-emission by optically-thick matter within the innermost stable circular orbit.
\end{abstract}
\begin{keywords}
accretion, accretion disks -- black hole physics -- galaxies: active -- X-rays: binaries.
\end{keywords}

\section{Introduction}
\label{intro}

In recent years, X-ray emission in the lamppost geometry \citep{mm96,mf04,dauser10,dauser16,garcia14} has become a popular model for accreting black holes (BHs) for both binaries containing a BH and Seyfert galaxies, e.g., \citet{parker14,parker15,keck15,beuchert17,basak17,xu18,tomsick18}. Also, it has been applied to accreting neutron stars, e.g., \citet{degenaar15}. In it, a point-like X-ray source is located on the BH rotation axis, which is perpendicular to a surrounding flat disk. The height of the lamppost, $h$, and the disc inner truncation radius, $r_\mathrm{in}$, are free parameters of the model, as well the BH dimensionless angular momentum, $a$. The BH horizon is at $r_\mathrm{hor}=[1+(1-a^2)]^{1/2}$, and the innermost stable circular orbit (ISCO) is at $r_\mathrm{ISCO}(a)$ derived by \citet{bardeen72}. Hereafter $r_{\rm hor}$, $r_\mathrm{ISCO}$, $h$ and $r_\mathrm{in}$ are expressed in units of the gravitational radius, $R_\mathrm{g}\equiv{GM/c^2}$, where $M$ is the BH mass. 

In our previous work \citep*{niedzwiecki16}, we pointed out the importance of the reduction of the observed luminosity from the lamppost due to photon trapping, time dilation and gravitational redshift, which effects have been neglected in most of lamppost studies. This reduction, severe for low lamppost heights, which are often found as best fits, may require a dramatic decrease of the accretion efficiency and the corresponding increase of the inferred mass accretion rate. We also pointed out the importance of e$^\pm$ pair production within the lamppost, which implies that many of fitted models are not physical, with the pair production rate greatly exceeding that of the annihilation. However, that work concentrated on the spectra from the lamppost/disc model, and we did not explore the parameter space for the flux attenuation, which we present in the present work. 

In this work, we point out in turn the importance of the radiation emitted by the lamp present on the side of the disc opposite to that on the side of the observer, which we hereafter refer as the bottom lamp. That emission is observable for $r_\mathrm{in} \ga 3$, and then it can increase the total emission for close to face-on observers by as much as an order of magnitude. This effect has been neglected in all previous studies of the lamppost geometry, as well as in papers fitting the lamppost model to X-ray data. In particular, this effect is not taken into account in the popular fitting code {\tt relxilllp} \citep{garcia14}. 

We also take into account the (usually neglected) radiation of the top lamp which is deflected back to the observer after crossing the equatorial plane twice, see, e.g., \citet{luminet79}. Also, we consider the emission of the bottom lamp circling around the BH on the side opposite to that of the observer. 

These effects can have a major effect on spectral fitting yielding truncated discs, e.g., in the hard state of BH binaries (e.g., \citealt*{done10,kdd14,plant15,bz16,basak17}). This then can be of major importance for the ongoing discussion about the geometry of that state, see, e.g., \citet*{dgk07}, \citet*{pvz18}.

Finally, we point out that the free-falling matter within the ISCO is optically thick (for a disc without truncation) even for moderate mass accretion rates. Then, $r_{\rm in}$ can be equal to $r_{\rm hor}$, and thus less than $r_{\rm ISCO}$, which is usually imposed as the minimum possible disc radius in fitting routines.

We present the method of calculations in Section \ref{method}. Selected results showing the importance of the included effects are given in Section \ref{results}. We summarize our main conclusions in Section \ref{conclusions}.

\section{The method and assumptions}
\label{method}

In our calculations, we use the GR model of \citet{niedzwiecki08}, with which we calculate the observed luminosity of both of the lamps and of the disc. In addition, we consider material free-falling within $r_\mathrm{ISCO}$, with the constants of motion of the ISCO \citep{cunningham75}. We note that such a material remains optically thick already for relatively moderate mass accretion rates of $\dot M \ga L_\mathrm{E}/c^2$ (where $L_\mathrm{E}$ is the Eddington luminosity) even for nonrotating black holes \citep{reynolds97}. Therefore, we hereafter assume that the optically-thick disc extends all the way to the horizon in cases without truncation. 

We neglect internal dissipation in the disc and we assume that only the external irradiation (from the lamppost) gives rise to the disc emission. For the disc emission, we assume that the total irradiating energy flux is locally re-emitted with an isotropic distribution with a constant specific intensity, i.e., the emitted flux in the disc rest frame is $F(\theta_\mathrm{em}) \propto \cos \theta_\mathrm{em}$, where $\theta_\mathrm{em}$ is the emission angle with respect to the direction normal to the disc plane. Due to energy conservation, 100 per cent of the flux incident on an optically-thick disc (\citealt{ss73}) is locally re-emitted. This re-emission consists of three components, namely that from Compton reflection of the incident photons (dominant at hard X-rays), a modified blackbody due to thermalization of the absorbed photons within the disc (dominant at UV/soft X-rays), and many spectral lines across the spectrum (in particular the Fe K$\alpha$ line at 6.4--7.0 keV in the rest frame), see, e.g., \citet{garcia10}. Here, we consider only the total re-emission, i.e., the sum of these components. Secondary re-emission of disc radiation which returns to the disc \citep{cunningham76} is also taken into account.
 
\begin{figure*}
\centering
\includegraphics[width=10.cm]{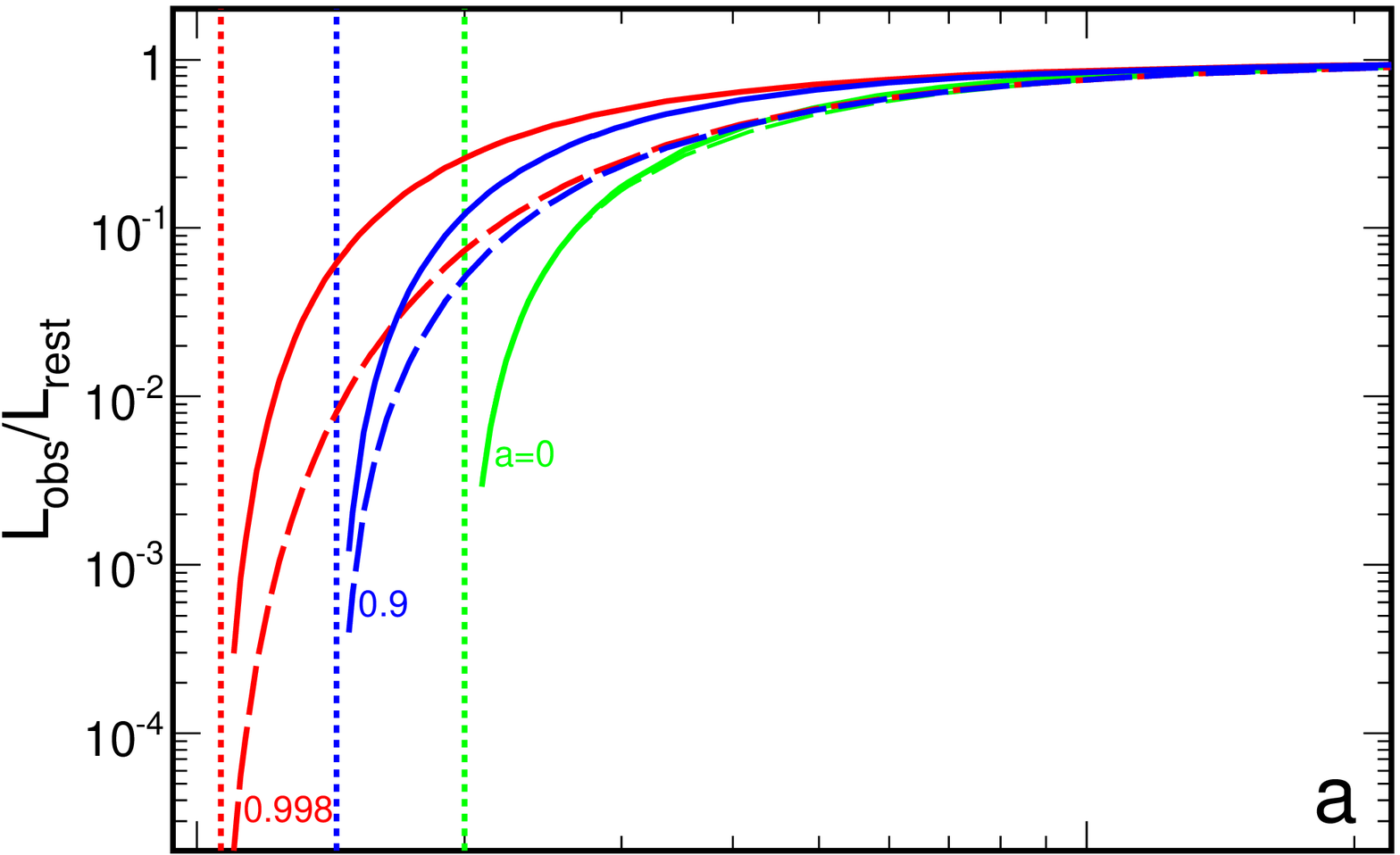}
\includegraphics[width=10.cm]{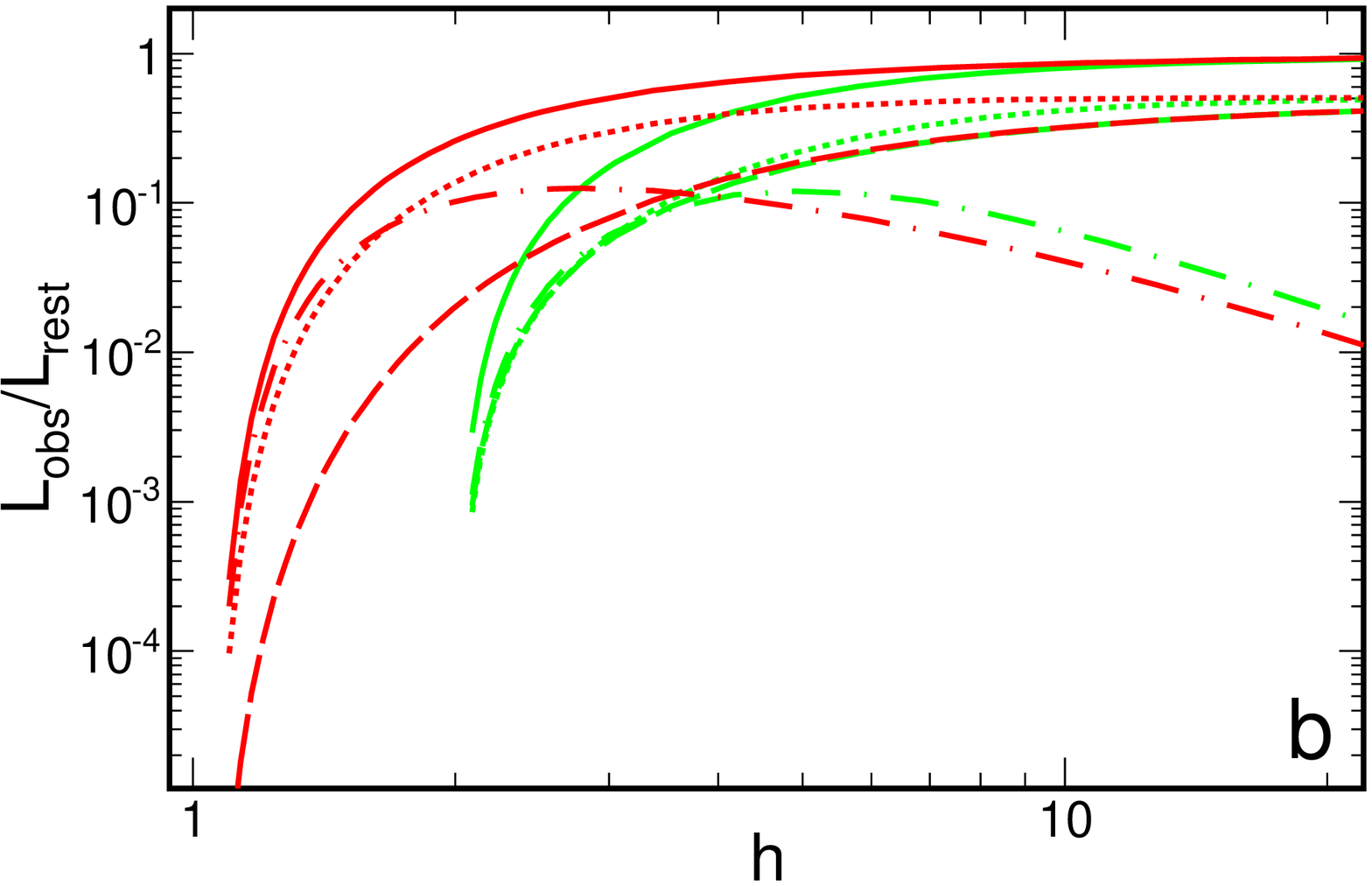}
\caption{The observed angle-integrated luminosity, $L_\mathrm{obs}$, vs.\ the lamp height. (a) The cases without a disc and with an untruncated disc. The dashed curves show $L_\mathrm{obs}$ of two lamps located at $h$ without an optically thick disc. The solid curves show $L_\mathrm{obs}$ of the top lamp and the disc with $r_\mathrm{in} = r_\mathrm{hor}$, including the second order re-emission. The red, blue and green curves are for $a=0.998$, 0.9 and 0, respectively. The dotted lines show the location of the event horizon. (b) The contributions of the disc and the lamp for the case with an untruncated disc, illustrating the effects of the 2nd order disc emission for $a=0.998$ and of the emission from $r < r_\mathrm{ISCO}$ for $a=0$. The solid curves show $L_\mathrm{obs}$ of the top lamp with the disc for $a=0.998$ and 0, same as in panel (a). The dashed curves show the contribution of the direct radiation from the lamp. For $a=0.998$, the red dot-dashed curve shows the luminosity observed due to the 2nd order re-emission of the returning disc radiation; the dotted red curve show luminosity from the re-emission of the primary emission only. For $a=0$, the green dot-dashed and dotted curves show the luminosity of the disc material below and above $r_\mathrm{ISCO}$, respectively.}
\label{hspin}
\end{figure*}

\begin{figure}
\centering
\includegraphics[width=\columnwidth]{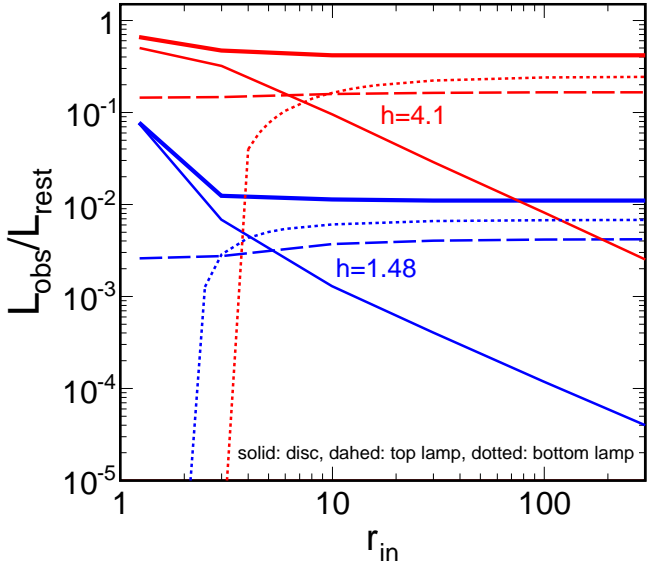}
\caption{The total, angle-integrated, observed luminosity (thicker solid curves) and its components as a function of $r_\mathrm{in}$ for $h=1.48$ (blue) and $h=4.1$ (red); $a=0.998$. The dotted curves are for the bottom lamp, the dashes curves are for the top lamp and the thinner solid curves are for the disc including the 2nd order emission.}
\label{rin}
\end{figure}

\begin{figure}
\centering
\includegraphics[width=\columnwidth]{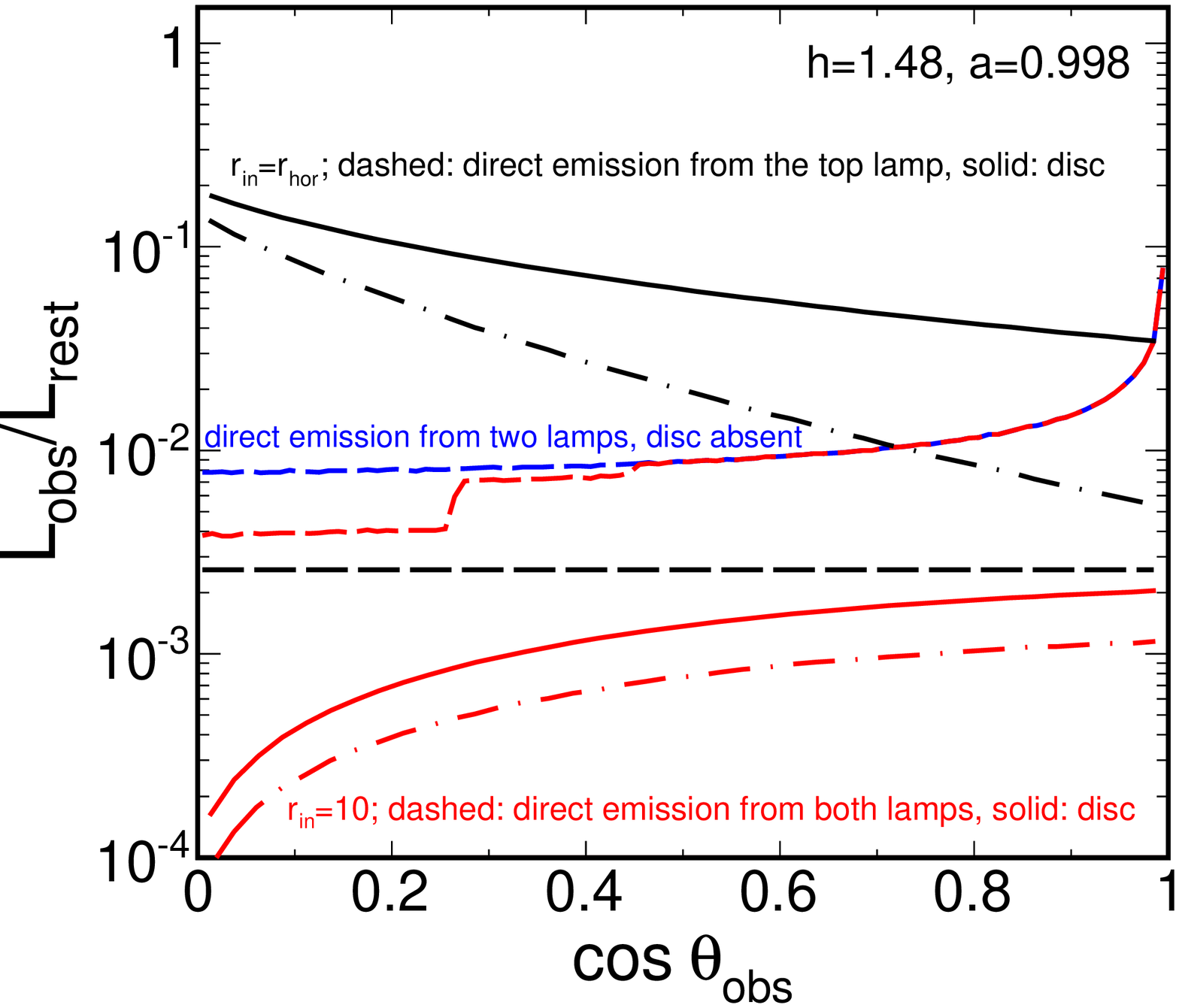}
\caption{Dependence of components of the observed radiation on the observation angle for $h=1.48$ and $a=0.998$. The blue dashed curve shows the luminosity of two lamps when no optically thick disc is present. The red curves are for two lamps and disc with $r_\mathrm{in} = 10$; the solid curve shows the luminosity of the disc including irradiation by both lamps; the dot-dashed curve is for the disc including only irradiation by the top lamp; the dashed curve is for the two lamps. The black curves are for the lamp above the disc with $r_\mathrm{in} = r_\mathrm{hor}$; the dashed curve is for the top lamp (the bottom lamp is not observable); the solid curve is for the disc including the 2nd order emission; the dot-dashed curve is for the disc neglecting the 2nd order emission.
}
\label{h148}
\end{figure}

\section{Results}
\label{results}

We have performed a number of calculations of the dependence of the ratio of the observed luminosity, $L_{\rm obs}$, to that intrinsically emitted by the lamppost, $L_{\rm rest}$, on the parameters $r_{\rm in}$, $a$, $h$ and the viewing angle at infinity, $\theta_{\rm obs}$. Note that $L_{\rm rest}$ equals twice the luminosity of a single lamp. We first note a few effects, which are crucial for the bolometric luminosity at infinity of the lamps/disc system.

1. If no optically thick disc is present in the vicinity of the BH, all photon trajectories originating from the symmetry axis and crossing the equatorial plane at $r < r_\mathrm{trap}$ are trapped under the event horizon, with the trapping radius of $r_\mathrm{trap} \la 3$, which only weakly depends on $h$ and $a$. E.g., for $a=0.998$, $r_\mathrm{trap} \simeq 2.5$ for $h=1.5$, and $r_\mathrm{trap} \simeq 3.5$ for $h=4$.

2. In the vicinity of rapidly rotating BHs, the reduction of the photon flux reaching the observer due to photon trapping for the disc re-radiation can be much weaker than the analogous reduction of the direct flux of the lamp. E.g., for $a=0.998$, $n_\mathrm{esc} \simeq 0.04$ for a static source at $h=1.5$ (a lamppost) and $n_\mathrm{esc} \simeq 0.28$ for a source on the Keplerian orbit at $r=1.5$ (a ring of the disc). Here $n_{\rm esc}$ is defined as the fraction of photons emitted by either the lamppost or a Keplerian source in the disc midplane that escapes to infinity. This effect takes place only if the infall velocity, $v^r$, is not large, which requires $r$ to be not much below $r_\mathrm{ISCO}(a)$. E.g., $n_\mathrm{esc} \simeq 0.01$ for $a=0.9$ at $r=1.5$, where $v^r \simeq -0.5 c$.

3. The infall velocity within $r_\mathrm{ISCO}$, and hence the collimation toward the BH horizon, increase with decreasing $a$. Therefore, weaker contribution of disc radiation from the innermost few gravitational radii can be expected for lower spin values, even if large optical thickness down to $r_\mathrm{hor}$ is assumed.

We present here our selected results. Fig.\ \ref{hspin}(a) shows $L_{\rm obs}/L_{\rm rest}$ as a function of the lamppost height, $h$ for three values of the dimensionless BH spin illustrating the difference between the cases with no surrounding disc (dashed curves) and one with an untruncated disc (solid curves). As implied by the item 2 above, the latter always yields a higher $L_{\rm obs}$. For given $a$ and $h$, $L_\mathrm{obs}$ depends on $r_\mathrm{in}$ and it has values in the range constrained by dashed and solid curves. For high BH spin values and $h \la 5$, the luminosity observed when the disc extends down to $r_\mathrm{hor}$ is substantially larger than that from a system with a truncated disc. This is due to re-emission of radiation, which would be trapped without the disc at $r \la 3$. For high BH spin values and $h \ga 5$, the effect of irradiation and re-emission at $r < 3$ is minor and the bolometric luminosity is almost independent of $r_\mathrm{in}$. On the other hand, for low spin values, radiation re-emitted at $r \la 3$ (i.e., by the optically-thick matter within $r_{\rm ISCO}$) is collimated toward the BH and the presence of an optically thick material leads to a minor increase of the observed luminosity, see the item 3 above. 

Fig.\ \ref{hspin}(b) shows components of $L_{\rm obs}/L_{\rm rest}$ for the case with an untruncated disc, and for $a=0$ and 0.998. The dashed curves show the contribution of the direct radiation (that not hitting the disc). We see it is a tiny fraction of $L_{\rm rest}$ at low $h$ and high spins, as well as it is then much less than the luminosity from the disc re-emission. For $a=0.998$, the dotted and dot-dashed curves show the observed contribution of the 1st and 2nd order of the re-emission, i.e., the latter shows the contribution from radiation that returned to the disc and was re-emitted. We see the 2nd-order emission is higher than the first order for $h\la 1.5$. Generally, the 2nd-order re-emission is important only for $a \ga 0.9$, for small $h$, and when the disc is untruncated.

We have checked that the orders of the re-emission higher than two have only minor importance. We have found that the 3rd order has the luminosity of $\la$30 per cent of the 2nd order. Taking it into account would increase the total disc luminosity by the maximum of $\sim$20 per cent (at low $h$) and $\la$10 per cent for $h>2$. 

For $a=0$, the dotted and dot-dashed curves in Fig.\ \ref{hspin}(b) show the contributions from re-emission below and above $r_{\rm ISCO}$, respectively. We see that the two components are comparable for $h\la 3$. Generally, contribution of radiation re-emitted within $r_\mathrm{ISCO}$ is important only for $h \la r_\mathrm{ISCO}(a)$; for $a=0$, it increases the total luminosity of the disc component by a factor of $\sim 2$.

Figs.\ \ref{rin}--\ref{h148} illustrate the importance of the emission of the bottom lamp and its dependence on the disc truncation radius. The bottom lamp is observable for $r_\mathrm{in} \ga 3$. Its radiation is gravitationally bent along the BH rotation axis, i.e.\ toward observers at $\cos \theta_\mathrm{obs} \sim 1$. Then, surprisingly, its contribution to the angle-averaged $L_\mathrm{obs}$ can be stronger than that of the top lamp, compare the dotted curves (for the bottom lamp) with the dashed curves (for the top lamp) in Fig.\ \ref{rin}. This is an effect of the amplification by the BH acting as a lens. For $h=1.48$, it happens for $r_{\rm in}\ga 3$, and for $h=4.1$, for $r_{\rm in}\ga 9$. Fig.\ \ref{rin} also shows that the disc re-emission dominates over the direct emission from both lamps for $r_{\rm in}\la 3$--4.

\begin{figure}
\centering
\includegraphics[width=8cm]{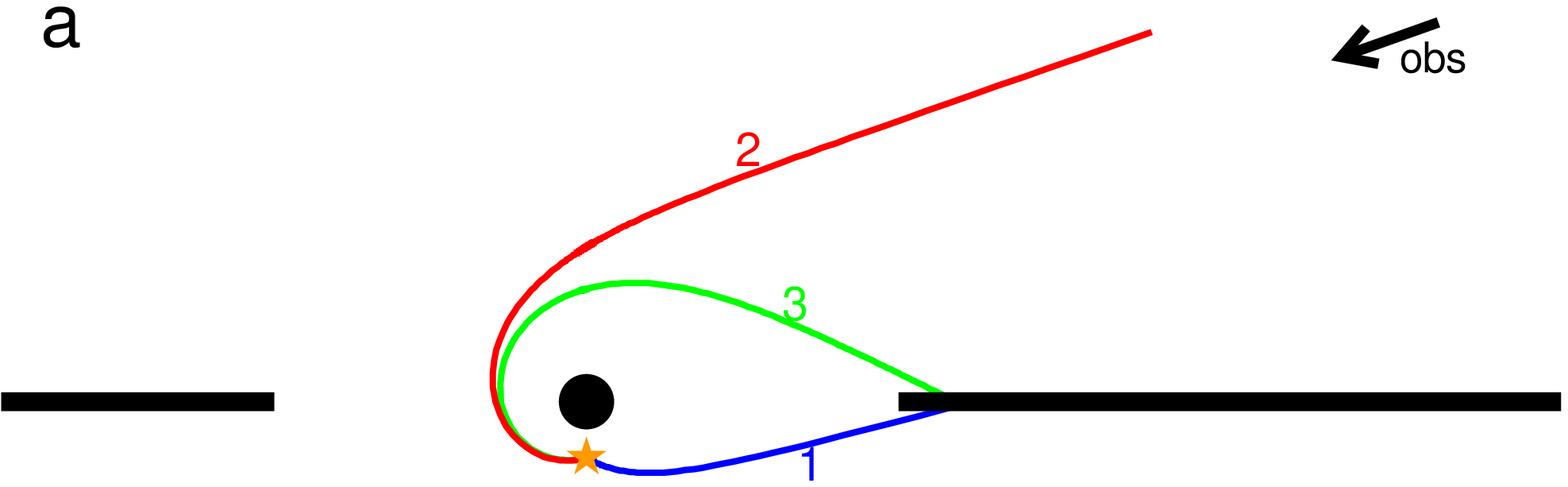}
\includegraphics[width=8cm]{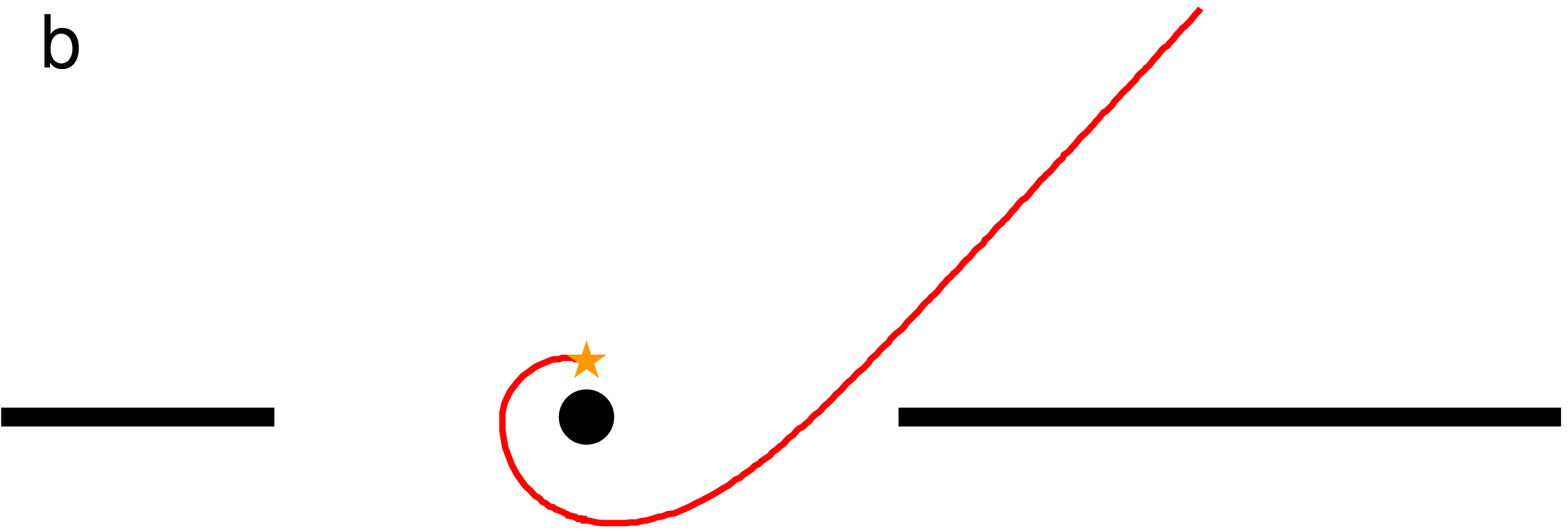}
\caption{(a) An illustration of the radiation of the bottom lamp going around the BH and either irradiating the disc (path 3) or reaching the observer (path 2) even when the direct path (path 1) to the observer is obscured by the disc. (b) Radiation from the top lamp deflected back by the BH.}
\label{bottom_lamp}
\end{figure}

For $r_{\rm in}\ga 3$, radiation of the bottom lamp crossing the equatorial plane contributes also to the disc irradiation, see Fig.\ \ref{bottom_lamp}(a). Its radial profile of irradiation is flatter than that of the top lamp and for small $h$ both lamps give similar contributions to the disc re-emision at $r \ga 10$, compare the solid and dot-dashed red curves in Fig.\ \ref{h148}.

Similarly to the emission of the bottom lamp, radiation from the top lamp deflected back by the BH (i.e., crossing twice the equatorial plane, see Fig.\ \ref{bottom_lamp}(b)) can be observed only for $r_\mathrm{in} \ga 3$. It does have a notable effect for small $h$, e.g.\ for $h=1.48$ it increases the observed luminosity of the top lamp by $\sim$50 per cent, as illustrated by the blue dashed curve in Fig.\ \ref{rin} increasing with the increasing $r_\mathrm{in}$, while that emission would be constant without that effect.

Fig.\ \ref{h148} shows some cases of the angular dependence of the observed radiation. It illustrates the effect of radiation from the bottom lamp propagating in front of the BH (with respect to the observer) being obscured by the disc for observation angles larger than a certain value, dependent on both $h$ and $r_\mathrm{in}$, see Fig.\ \ref{bottom_lamp}(a). E.g., for $h=1.48$ and $r_\mathrm{in} = 10$, it is obscured at $\cos \theta_\mathrm{obs} \la 0.25$, as shown by the drop of the red dashed curve at this $\theta_\mathrm{obs}$. However, this drop is to a level that still includes radiation from the bottom lamp going to the observer behind the BH, see Fig.\ \ref{bottom_lamp}(a), as shown by the difference between the black and red dashed curves at $\cos \theta_\mathrm{obs} < 0.25$. A weaker drop at $\cos \theta_\mathrm{obs} \simeq 0.45$ is due to obscuration of the back-deflected radiation of the top lamp. 

Fig.\ \ref{rin} shows that for $r_\mathrm{in} \ga 3$, the angle-averaged $L_\mathrm{obs}$ only weakly depends on the value of $r_\mathrm{in}$. However, a change of $r_\mathrm{in}$ changes the relative contributions of the direct and disc radiation and thus it affects both the angular distribution and the spectrum at infinity. At a high spin, the disc radiation emitted from $r \la 3$ is gravitationally and kinematically beamed along the equatorial plane. Then, for small $h$, the 1st-order re-emission is directed mostly edge-on, as shown by the dot-dashed black curve in Fig.\ \ref{h148}. However, the radial redistribution of the irradiating flux by the returning radiation makes the dependence on $\theta_\mathrm{obs}$ weaker, as shown by the solid black curve in Fig.\ \ref{h148}, because the 2nd-order re-emission arises on average at larger distances than the 1st-order one. 

An increase of $r_\mathrm{in}$ above $\simeq 3$ can result in a substantial increase of the direct flux, especially for face-on observers. This is illustrated for $h = 1.48$ in Fig.\ \ref{h148}, where we see that the direct radiation from the lamps observed at $\theta_\mathrm{obs} \la 20 \degr$ increases by over an order of magnitude. This increase is due to the lensed contribution of the bottom lamp. For static lamps, the observed spectra of from both lamps are the same, and thus this rapid brightening occurs without the change of the spectrum.

\section{Conclusions}
\label{conclusions}

We have performed a systematic study of the BH lamppost model taking into account the presence of the bottom lamp, which was neglected in its previous studies. We have calculated and presented the dependence of the angle-averaged attenuation of the intrinsically-generated radiation as functions of the lamppost height, the BH spin and disc truncation. As also stressed in our previous work \citep{niedzwiecki16}, that attenuation is very severe, by orders of magnitude, for low lamppost heights, which heights are often found when fitting the X-ray data (e.g., \citealt{keck15,degenaar15,parker15}). 

We then take into account, for the first time, the emission of the bottom lamp. Its angle-integrated emission at infinity is larger than that of the top lamp for a truncated disc, which effect is due to gravitational lensing. In particular, the observed flux of the bottom lamp can exceed that of the top one by more than an order of magnitude for low viewing angles even when the truncation radius is small. 

We also show that the effects of the emission of the top lamp circling the BH and crossing the equatorial plane twice and that of the emission of the bottom lamp circling the black hole on the side opposite to that of the observer are observable, and can substantially affect the observed fluxes. 

A major observational importance of effects considered in this work concerns their impact on the reflection strength, i.e., the relative normalization of the primary and reflected components. Fitting the relativistic reflection can give tight constraints on both the disc truncation radius and the observer inclination, which determine the expected amount of the latter component. Then, contribution from the bottom lamp can significantly reduce the reflection strength. Thus, the considered effects should be taken into account when fitting X-ray data of accreting BHs. They can be used, in particular, to test the self-consistency of the lamppost model, and provide a useful diagnostic to resolve the currently ongoing dispute about the nature of the hard state of BH binaries.

\section*{ACKNOWLEDGMENTS}

We thank Thomas Dauser and Joern Wilms for discussions. This research has been supported in part by the Polish National Science Centre grants 2013/10/M/ST9/00729, 2015/18/A/ST9/00746 and 2016/21/B/ST9/02388.

\label{lastpage}
\end{document}